\documentclass[epj]{svjour}
\usepackage{graphics}

\begin{document}
\title{A scheme for the preparation of  a polarised antineutron beam
     }
\author{Berthold Schoch%
}                     
\institute{Physikalisches Institut, Universit\"at Bonn, Nussallee 12, D-53115 Bonn, Germany }
\date{Received: date / Revised version: date}
%
\abstract{A polarised antineutron beam can be prepared via 
electro production in the reaction  $\gamma +\overline{p}\rightarrow \pi ^{-}+%
\overline{n}$ by using circularly polarised virtual photons as projectils
hitting a stored antiproton beam in a storage ring. High luminosities can be
achieved due to the progress in cooling techniques for the antiproton beam
and the development of polarised electron beams with high current, high
polarisation and low emittance. In addition, multiple interaction points can
be used. A tagged antineutron beam with intensities  of the order of $N_{%
\overline{n}}=10^{3}-10^{5}$ (s$^{-1})$ seems to be achievable. A
concentration of intensity resides in the momentum range $18$ $\leq p_{%
\overline{n}}$ $(GeV/c)$ $\leq $ $19$ for a stored antiproton beam with a
momentum of $20$ $GeV/c.$ The total momentum range is given by $14$ $\leq p_{%
\overline{n}}$ $(GeV/c)$ $\leq $ $19$. A cone of 7 mrad contains all
produced antineutrons. A polarisation of $P_{%
\overline{n}}$ = 88\% of the antineutron beam can be achieved. }

%
\maketitle
\section{Introduction}
\label{sec:1}
\vspace{1pt}
Further progress in hadron physics depends critically on possibilities
to carry out experimental
programs using polarised beams and targets.
In a review article \cite{Ref1} the authors state: "The difficulties in
obtaining beams of antineutrons of suitable intensity and energy definition
were overwhelming". 
The preparation of polarised
antineutron beams is especially challenging.
So far, only few experiments with unpolarised antineutrons as projectils 
have been carried out. At the AGS accelerator, Brookhaven, USA, 
an antineutron beam has been used with momenta up to
500 MeV/c \cite{Ref2}. An antineutron beam with momenta up to 400 MeV/c 
has been prepared at the CERN laboratory, Geneva, Switzerland, for experiments with 
the OBELIX detector \cite{Ref3}. 
Thereby, antineutron intensities of the order of $N_{\overline{n}}=10^{2}/s$
at reaction target have been obtained via the charge exchange reaction
p+$\overline{p}$  $\rightarrow$n+$\overline{n}$.
A tagged antineutron beam in the momentum range 5 $\leq p_{\overline{n}}$ (GeV/c))$\leq 85$
has been made available by the upgraded MIPP-spectrometer at the FERMI laboratory,
Batavia, USA, providing an
antineutron flux of $N_{\overline{n}}=10^{-1}/s$ via the reaction 
p+$\overline{p}$ $\rightarrow$  p+$\pi^{-}$+$\overline{n}$ \cite{Ref4}.
Polarised antineutron beams have not been available, so far,
for physics experiments. 
\newline
A method described in \cite{Ref5} makes use of the different absorption cross
sections, $\sigma _{3/2}$ and $\sigma _{1/2}$ for the
two spin projections of antiprotons stored in a storage ring interacting
with circularly polarised $\gamma -$ray radiation in order to polarise antiprotons and,
as a byproduct, produce polarised antineutrons.
Thereby, polarised antineutrons with the same helicity will be produced
by both spin projections $J_{1/2} $ and $J_{3/2}$ provided that the
$\gamma$-energy stays below the two pion production threshold.
In the $J_{1/2}$-channel the $\pi ^{-} $ production on the antiproton leads to a spin flip
of the antineutron, in the $J_{3/2}$-channel, however, $\pi ^{-} $ production does not
change the spin of the participating antinucleon.  
\begin{figure}{}
\resizebox{0.50\textwidth}{!}{%
  \includegraphics{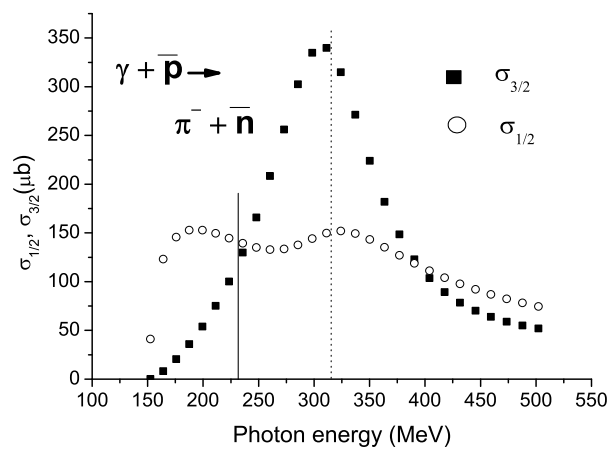}
}
\caption{The absorption cross sections $\sigma _{1/2}$ and
$\sigma _{3/2}$ for $\gamma$+$\overline{p}$
$\longrightarrow {\pi}^{-}$ + $\overline{n}$ calculated with the MAID program \protect\cite{Ref6}, 
upper limit of integration (vertical line), two pion production threshold (dotted line)}
\label{fig:1}       
\end{figure}
That characteristics of the reaction opens the door for a creation of an antineutron
beam with high polarisation.
Fig. 1 shows the cross sections $\sigma _{1/2}$ and
$\sigma _{3/2}$  for the reaction $\gamma$+$\overline{p}$
$\longrightarrow {\pi}^{-}$ + $\overline{n}$.
\newline 
Especially, 
the availability of a highly polarised antineutron beam in a 
range of energies around 10-20 GeV and
intensities of around $N_{\overline{n}}=10^{3-5}/s$ will open up
new classes of experiments in hadron physics. Together with a polarised antiproton beam 
complete experiments in the
spin and isospin
space can be performed and the time like region of the momentum transfer be explored
\textit{e.g.} with dileptons in the final state.
For a possible application of the method the major question to answer  will be: can 
a luminosity be reached to prepare
a polarised antineutron beam with the intensities addressed above,
preferentially, without 
the need of an antiproton ring with extra large acceptances.
 
\section{Electro production versus photo production }
\label{sec:2}
 
The reactions $\gamma +\overline{p}\rightarrow \pi ^{-}+\overline{n}$ and $%
\gamma +\overline{p}\rightarrow \pi ^{0}+\overline{p}$ can be realised by
real and virtual photons.
An electron beam instead of  a real photon beam can be used to extract from electro induced
reactions photon cross sections. Thereby, the concept
of virtual photons provides the theoretical framework for that application \cite{Ref7}.
By considering virtual photon energies close to the incoming electron
energies,
the so called end point region, the longitudinal contributions to the
reaction cross sections are very small. By integrating over the angle 
$\theta _{e^{\prime }}$ 
of the scattered electrons and applying the forward peaking approximation,
$\theta _{e^{\prime }}=0,$ for evaluating the response of the hadronic
structure, the electro- and
photo production cross sections can be related. That concept of virtual
photons has been
checked experimentally to an accuracy of a few percent \cite{Ref8}.
The spectrum of virtual photons as a function of photon energy resembles, at
the end point energy region, to a bremsstrahl spectrum. The intensity of virtual
photons
corresponds to a bremsstrahl with a radiator of the order of one percent of a radiation
length.
One of the big advantages by using an electron beam, compared to a $\gamma$-beam, resides
in the possibility to use rather
conveniently a large number of interaction points between electron and antiproton beam.

\section{Ingredients determining the luminosity }
\label{sec:3}

The luminosity of an external electron beam interacting head on with a stored antiproton beam can be written as
\begin{equation}
L=N_{\overline{p}}\cdot \frac{N_{e }}{4\pi \cdot \sigma _{x}\cdot
\sigma _{y}}\cdot \nu _{c}
\end{equation}
with the number of stored antiprotons $ N_{\overline{p}}$
hitting a target of electrons of
density $\frac{N_{e }}{4\pi \cdot \sigma _{x}\cdot \sigma _{y}}$
(electrons/cm$^{2} \cdot s$) and with the frequency of the antiproton bunch of $\nu _{c}$.
$\sigma _{x}$ and $\sigma _{y}$
stand for the Gaussian beam profiles of the 
two intersecting beams of same extension.
By choosing a virtual $\gamma$-energy range of $E_{\gamma
}^{threshold}\leq E_{\gamma } \leq 230$ MeV in the rest system of the antiproton,
the range of the charged pion production channel 
$\gamma +\overline{p}\rightarrow \pi ^{-}+\overline{n}$
can be covered in that system as indicated in fig.1. by the vertical line 
as the upper limit. 
An electron energy of 230 MeV in the rest system of the antiproton transforms into  
$E_{e}$=5.39 MeV in the laboratory system by assuming an antiproton momentum
in the storage ring of 20 GeV/c.
\newline
In order to use realistic numbers for the calculation of the luminosity
values of beam parameters for a stored antiproton beam are taken as have been reported
in the design report for the planned facility FAIR at GSI \cite{Ref9},\cite{Ref10},
Darmstadt, Germany, foreseen for its High Energy Storage Ring (HESR)
for the storage of antiprotons. The FAIR facility provides up
to $n_{\overline{p}}=7\cdot 10^{10}/h$ antiprotons, thereby, 
the space charge limit $N_{\overline{p}%
}^{limit}$ for the number of stored antiprotons for a ring like HESR is
$N_{\overline{p}}^{\textit{limit}}\succeq 10^{13}.$  Beam diameters $d_{\overline{p}
}^{beam}$ at the interaction region of $d_{\overline{p}}^{beam}=10 \mu m$ can
be achieved.
\newline
A tremendous progress has been made over the past fifteen years by
preparing intense polarised electron beams  with excellent emittances
and high polarisation, see \cite{Ref5}. 
Several groups are working on a further 
improvement of the emittance of the beam by using \textit{e.g.} cw guns. 
An increase of the current by a factor of up to 100 and an improvement of the emittance
by a factor of 5-10 are envisaged in order to
meet the needs for proposed Electron Ion Colliders (EIC). In addition, on
the way to develop a fourth generation of synchrotron radiation sources the
mode of running electron linear accelerators in an energy recovering mode has been
tested successfully \cite{Ref11},\cite{Ref12}. Applications of that running mode, besides for synchrotron
radiation, have not been reported so far but might be the option 
for the future electron antiproton interactions.  
With those ingredients the luminosity of the electron antiproton
interactions can be calculated.
\newline
The following values are, finally, used for the calculation of the luminosity:
$N_{\overline{p}}$=2 $\cdot 10^{12}$, that means a filling time of the antiproton ring
of 28h, $\nu _{c}$=6 $\cdot 10^{5}(s_{-1})$, assuming a length of the antiproton ring of 500 m,
$N_{e}$=6 $\cdot 10^{15} $, corresponding to an electron current of 1 mA 
and, finally, $\sigma _{x,y}$=$10\mu m$.
With those values the luminosity for one interaction point (I.P.) yields by using eq. 1:
$L_{e}^{\textit{one I.P.}}$=9.55 $ \cdot 10^{32}$ ($cm^{-2}s^{-1}$).
\newline
Depending on the lattice of the antiproton storage ring 
various scenarios of the e-$\overline{p}$ interaction region
can be considered. The design of the lattice of the antiproton ring
determines the diameter of the antiproton 
beam at the interaction region
as well as a possible extension of that region. Such a properly chosen extension allows
to implement tools like quadrupole triplets, dipols formed as wigglers or 
a solenoid in order to achieve more interaction points \cite{Ref13}.
The solenoid will be chosen as an example.  
A beam with low energy electrons of $E_{e}$=5.39 MeV in the laboratory  
traversing through a solenoid with a high longitudinally magnetic field
creates many foci per length. That field has almost no influence on an
antiproton beam with a momentum of 20 GeV/c.
With a magnetic field strength B of 10 T, \textit{e.g.} \cite{Ref14}, 88 foci per m can be reached.  
Thus many interaction points can be created on a relatively short length 
by implementing a strong longitudinal magnetic field.
Using an interaction zone with a magnetic field strength of
10 T and an extension of 1 m a luminosity of $L_{e}$=8.4 $ \cdot 10^{34}$ ($cm^{-2}s^{-1}$)
could be achieved.
That luminosity is used to calculate reaction rates. 

\section{Reaction rates
}
\label{sec:4}
With the luminosity $L_{e}$ and the cross sections shown fig.1 
the reaction rates for the different channels can be calculated.
Each antiproton spin channel will be treated separately with a luminosity $%
L_{1/2,3/2}^{e-\overline{p}}=\frac{L_{e}}{2}$ taking into
account that each channel contains half of the stored antiprotons. For each
channel the reaction rates contain contributions from the unpolarized part of the virtual
photon spectrum and the unpolarized part of the electron beam. The reaction
rates for the two spin components of the antiprotons $N_{1/2}$ and $N_{3/2}$
as well as the rates  of antineutrons $N_{unpol}^{virt}$ and \ $%
N_{e}^{unpol}$ due to the unpolarized part of the virtual photons
and the unpolarized part of the electron beam, respectively, are given by 
\newline
\begin{equation}
N=L\cdot \int_{threshold}^{E_{\gamma }^{\max }}x\cdot (h(E_{\gamma })\cdot 
\frac{dN_{\gamma }^{virtual}(E_{e})}{dE_{\gamma }}\cdot \sigma )dE_{\gamma }
\end{equation}%
with $N=N_{1/2,3/2}$, $L=L_{1/2,3/2}^{e-\overline{p}}$, $\sigma =\sigma
_{1/2,3/2}$ 
\begin{equation}
N=L\cdot \int_{threshold}^{E_{\gamma }^{\max }}x\cdot ((1-h(E_{\gamma
}))\cdot \frac{dN_{\gamma }^{virtual}(E_{e})}{dE_{\gamma }}\cdot \sigma
)dE_{\gamma }
\end{equation}%
with $N=N_{unpol}^{virt}$, $L=L_{e}$, $\sigma =\sigma
_{tot}$ 
\begin{equation}
N_{e}^{unpol}=L_{e}\cdot \int_{threshold}^{E_{\gamma
}^{\max }}((1-x)\cdot \frac{dN_{\gamma }^{virtual}(E_{e})}{dE_{\gamma }}%
\cdot \sigma _{tot})dE_{\gamma }
\end{equation}
$x$ stands for the degree of polarisation of the electrons.  $h(E_{\gamma })$
describes the helicity transfer from the electron beam to the virtual
photon, see eq. 5 \cite{Ref5}, $\frac{dN_{\gamma }(E_{e})}{dE_{\gamma }}$ represents 
the virtual photon spectrum and $\sigma _{tot}=\sigma _{1/2}+\sigma _{3/2}$.
Reaction rates are shown in table 1 for
the electron energy in the rest system of the antiproton $E_{e}=230 MeV$  
and $x=0.85.$ 
\bigskip
\newline
\begin{tabular}{||c||c||c||c||c||c||}
\hline\hline
$N_{1/2}$ & $N_{3/2}$ & $N_{unpol}^{virt}$ & $N_{e}^{unpol}$ & $%
N_{pol}^{total}$ & $N_{unpol}^{total}$ \\ \hline\hline
19814 & 5605 & 2998 & 5040 & 25419 & 8038 \\ \hline\hline
\end{tabular}
\bigskip
\newline
\textbf{Table 1} Production rates (antineutrons/s) of the different channels
\bigskip
\bigskip
\newline
The production rate of polarised antineutrons with polarised virtual photons  $N_{\overline{n}%
}(s^{-1})=N_{1/2}$ \ + $N_{3/2}$ amounts to $N_{\overline{n}%
}(s^{-1})$=25419 and the total rate of unpolarised antineutrons 
$N_{unpol}^{total}(s^{-1})=N_{unpol}^{virt}$ + $N_{e}^{unpol}$ 
=8038. Half of the unpolarised antineutrons have the same helicity as the polarised antineutrons thus
the polarisation of the antineutrons is given by
\begin{equation} 
P_{\overline{n}}=\frac{N_{\overline{n}}+\frac{N_{unpol}^{total}}{2}}{N_{%
\overline{n}}+N_{unpol}^{total}}
\end{equation}
 and yields $P_{%
\overline{n}}=88\%$  by using the rates given in table 1.
\section{ Properties of the antineutron beam
}
\label{5}
Besides the overall intensity and polarisation of the antineutron beam the
distribution of the longitudinal and transverse momenta are important in
order to judge its possible application for future experiments.
The distribution of the longitudinal momenta describes,  together with the
respective differential cross sections,
the concentration of the strength of momenta in the extracted antineutron
beam.
In fig. 2 the momenta for three $\gamma -$ energies 
are shown, $E_{\gamma }=160,190$ and $230$ MeV, as a function of the pion production angle $%
\theta _{\pi ^{-}}$  in the rest system of the
antiproton. 
\begin{figure}{}
\resizebox{0.50\textwidth}{!}{%
  \includegraphics{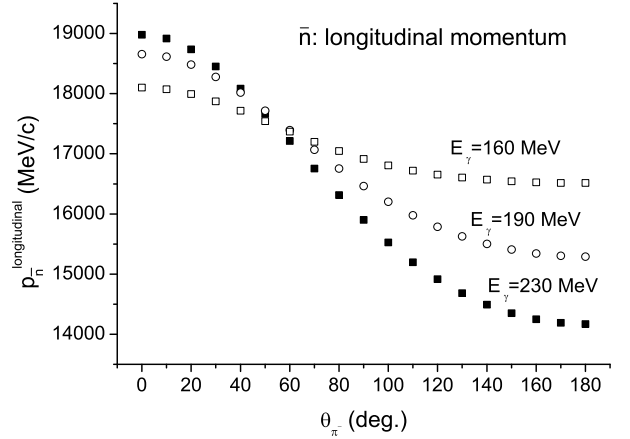}
}
\caption{Longitudinal momentum distributions of a beam of antineutrons in the 
laboratory system}
\label{fig:2}       
\end{figure}
A concentration of momenta can be seen in the momentum range 
18 GeV/c $\leq p_{\overline{n}} \leq$ 19 GeV/c. This range contains a
strength of around one third of the total  reaction cross section for the $%
\pi ^{-}$ - channel.
The distribution of the transverse momenta in the laboratory system determines 
the divergence of the beam and, thus, the range of applications in combination with suitable targets.
As in fig. 2 for the longitudinal momenta, in fig. 3 the transverse momenta are shown as a function of 
the pion production angle $
\theta _{\pi ^{-}}$ for three $\gamma -$ energies in the rest system of the
antiproton.
\begin{figure}{}
\resizebox{0.5\textwidth}{!}{%
  \includegraphics{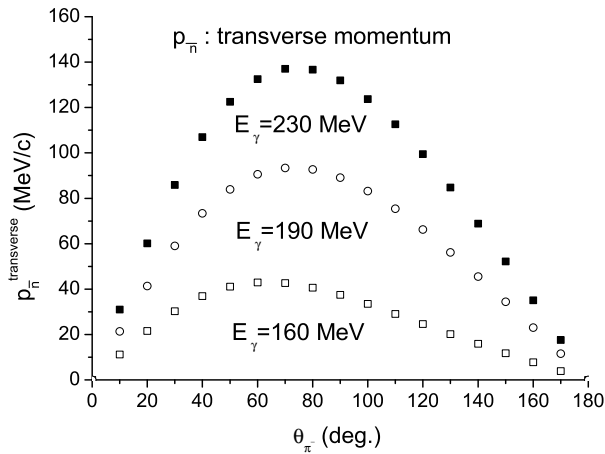}
}
\caption{Transverse momentum distributions of a beam of antineutrons in the 
laboratory system }
\label{fig:3}       
\end{figure}
The divergence of the antineutron beam is determined by the ratio $\frac{p_{%
\overline{n}}^{transverse}}{p_{\overline{n}}^{longitudinal}}$.
A  cone of 7 mrad contains all produced antineutrons. The antineutrons
concentrated in the narrow momentum range 18 GeV/c $\leq p_{\overline{n}}\leq$
19 GeV/c mentioned above show a divergence of around 5 mrad.\ 
As can be expected, the contents of fig(s) 1, 2 and 3 suggest that, provided
the luminosity can be reached, 
the ideal running mode would be close to the pion production threshold,
providing a narrow momentum range for all
produced antineutrons together with a small divergence of the beam which has been
one reason for choosing an electron energy of $E_{e}$=230 MeV in the rest system of the antiproton.
The
optimum running mode will also be
influenced by the divergence of the antiproton beam in the interaction zone
which might reach the same order
of divergence for the final antineutron beam.
\newline
The electro produced antineutrons are accompanied by negatively charged pions with
momenta in the few GeV/c region and low energy electrons. Those pions and electrons can be separated
from the stored antiproton beam. The
detection of  pion and/or electron provides a tag and thus a timing signal for antineutron induced
reactions in a reaction target. Those targets, unpolarised and polarised, might be placed 
into the antiproton beam line by providing a hole in the middle of the target for the antiproton beam.
With such a configuration luminosities up to L=$10^{28}cm^{-2}s^{-1}$ and more should become possible.
\section{Losses of antiprotons } 
\label{6}
Besides the reaction process $\gamma +\overline{p}\rightarrow \pi ^{-}+%
\overline{n}$ used to produce an antineutron beam two other reaction
channels have to be considered, namely $\gamma +\overline{p}\rightarrow \pi
^{0}+\overline{p}$ and $e+\overline{p}\rightarrow e^{\prime }+\overline{p}$
which might lead to losses of the stored antiproton beam. Those losses
depend strongly on the momentum acceptance of the storage ring. The
contribution of the  $\gamma +\overline{p}\rightarrow \pi ^{0}+\overline{p}$
channel remains up to $E_{\gamma }=230$ MeV below the rates of the $\gamma +%
\overline{p}\rightarrow \pi ^{-}+\overline{n}$ channel. The $e+\overline{p}%
\rightarrow e^{\prime }+\overline{p}$ channel, however, needs special attention. 
Via elastic electron scattering large momenta $\overrightarrow{q}$ can be
transferred from the electron to the antiproton. The cross section $\sigma
^{R}$ for this process is given by the Rosenbluth formula \cite{Ref15} and decreases for
low momentum transfers q at least with $\frac{1}{q^{4}}$ and is dominated by
the transverse transfer.
The rate of losses of stored antiprotons due to elastic electron scattering
is given by $N_{\overline{p}}^{loss}=L_{e}\cdot \int_{\theta _{acc}}^{\pi }%
\frac{d\sigma ^{R}}{d\Omega _{e^{\prime }}}d\Omega _{e^{\prime }}.$
Thereby, $\theta _{acc}$ stands for the electron scattering angle determined
by the momentum acceptance of the antiproton storage ring and $\frac{d\sigma
^{R}}{d\Omega _{e^{\prime }}}$ the differential cross section calculated via
the Rosenbluth formula. The calculation is carried out in the rest system of
the antiproton. An electron scattering angle $\theta_ {acc}=0.44(rad)$ results from the calculation
by using a momentum acceptance $\frac{\Delta p}{p}=0.005$ for the antiproton storage ring 
as foreseen
for the HESR of the FAIR facility \cite{Ref9}. With the result of the integral   
$\int_{\theta _{acc}}^{\pi }\frac{d\sigma
^{R}}{d\Omega _{e^{\prime }}}d\Omega _{e^{\prime }}=\allowbreak
1.\,\allowbreak 87\times 10^{-29} (cm^{2})$ and the luminosity $L_{e}$
the loss rate of antiprotons in the ring yields $N_{\overline{p}}^{loss}=\allowbreak 1.\,\allowbreak
57\times 10^{6}(s^{-1}).$
Thus, the loss rate due to elastic electron scattering exceeds the total number of
produced antineutrons reported in sec. 4 by a factor $\sim 50.$ This
factor could be
reduced in principle to $\sim 1$ by using an antiproton storage ring in a synchroton mode
with a momentum acceptance of $\frac{\Delta p}{p}\sim 0.1$. However,
as long as the loss rates of antiprotons in the storage ring are smaller
than the potential fill-up rates as in the case of the FAIR project the losses
can be tolerated. For the discussed case the losses are 10\% of the
potential maximum fill-up rate.
\newline
Electron antiproton interactions exchanging small momenta, $0 \prec \theta_{e} \leq \theta_{acc}$, 
heat up the
stored antiproton beam. Thus, a cooling system must be in place for the
antiproton storage ring. Cooling times of 100 - 1000 (s) as planned
for the HESR ring at the FAIR facility via stochastic cooling are sufficient
to provide the necessary cooling. 
\section{Summary}
\label{7}
With the proposed method \cite{Ref5} tagged polarised
antineutron beams can be prepared for the first time.  Antineutron fluxes
can be reached, exceeding the intensities reported so far, by choosing
optimal parameter combinations in eq. 1 for the luminosity.  The method can 
provide a wide coverage of antineutron energies by choosing
suitable setups. 
The first test measurements for the optimization of the method will certainly
use a stored proton beam. As a result a highly polarised neutron beam can be prepared via
the reaction
$\gamma +p\rightarrow \pi ^{+}+n$.

\end{document}